\documentclass[%
reprint,
superscriptaddress,
 amsmath,amssymb,
 aps,
]{revtex4-1}

\usepackage{braket}
\usepackage{xcolor}
\usepackage{graphicx}
\usepackage{dcolumn}
\usepackage{bm}
\usepackage{gensymb}
\usepackage{amsmath}
\usepackage{amsfonts}
\usepackage{amssymb}
\usepackage{mathtools}
\usepackage{lipsum}
\usepackage{comment}

\begin{document}
\title{Real-time stokes polarimetry based on a polarisation camera}

\author{Mitchell A. Cox}

\affiliation{School of Electrical and Information Engineering, University of the Witwatersrand, Johannesburg, South Africa}
\author{Carmelo Rosales-Guzm\'an}
\affiliation{Centro de Investigaciones en \'Optica, A. C., Loma del Bosque 115, Col. Lomas del Campestre, 37150, Le\'on, Gto., M\'exico}

\email{mitchell.cox@wits.ac.za} 

\homepage{https://www.wits.ac.za/oclab} 


\begin{abstract}
This lab note introduces the ``Stokes Camera,'' a simple and novel experimental arrangement for real-time measurement of spatial amplitude and polarisation and thus spatially resolved Stokes parameters. It uses a polarisation sensitive camera and a fixed quarter-wave plate, providing a one-shot, digital solution for polarisation measurement that is only limited by the frame rate of the camera and the computation speed of the provided code. The note also provides background information on relevant polarisation theory and vector vortex beams, which are used as a demonstration of the device.
\end{abstract}
\maketitle
\section{Introduction}

Structured light research involves the use of optical modes with diverse spatial amplitude, phase, and polarization profiles. The creation and detection of such modes are fundamental to the field, and various experimental techniques (often with the goal of being simpler, more accurate or more sensitive) have been developed to accomplish these tasks \cite{Pinnell2020,Tidwell1990,Niziev2006,passilly_simple_2005,Mendoza-Hernandez2019,Marrucci2006,naidoo2016controlled,Devlin2017,Cox2022}. 

In the last decade or two, the use of liquid crystal spatial light modulators has transformed the field of structured light because of the ease and efficiency of which the spatial amplitude and phase can be modulated \cite{SPIEbook,Davis2000,Maurer2007,Moreno2012,Mitchell2017,Rosales2017,Rong2014,Liu2018}. The polarisation state of a laser beam is an essential property in various scientific and engineering applications, and control of this degree of freedom has opened many new avenues in structured light research in the form of so-called vector modes \cite{Zhan2009,Rosales2018Review,Yao-Li2020,Zhaobo2021,Rosales2021,Hu2021}, which are useful in optical trapping\cite{Yuanjietweezers2021,Michihata2009,Kozawa2010,Donato2012,Zhao2005}, microscopy\cite{backscattering2021,Hao2010,Torok2004,Segawa2014,Segawa2014b}, metrology\cite{Ndagano2017,Hu2019,Toppel2014,Berg-Johansen2015} and even optical communications \cite{Trichili2020,cox2020slt,Barreiro2008,DAmbrosio2016,Otte2020,Ndagano2018,Milione2015,Milione2015e,Wang2017,Liu2018a,Willner2018}

Control of the spatial polarisation of a beam is significantly more challenging but recent progress is in part due to renewed interest in digital micro-mirror devices (DMDs) \cite{Lerner2012,Ren2015,Mitchell2016,Gong2014,manthalkar2020,Zhaobo2020,Cox2021dmdmod}. This renewed interest in DMDs is likely due to the ease of which the polarisation degree of freedom can be used and manipulated \cite{rosales-guzman_polarisation-insensitive_2020,Hu2021Random,Scholes2019,Zhao2019,Hu2022}.

Stokes polarimetry is a well-established technique in optics that has been used for over a century to study the polarisation properties of light. The technique was first introduced by Sir George Stokes in 1851 in his study of the composition of light. Since then, it has become an essential tool in many fields of optics, including laser physics, spectroscopy, and microscopy, among others.

Stokes parameters can be measured in various ways but the simplest and most intuitive is called the rotating wave-plate method \cite{Toninelli:19}. This method is based on the principle of passing the laser beam through a wave-plate at a specific angle and rotating the wave-plate. As the wave-plate rotates, the polarisation state of the laser beam changes, and the intensity of the laser beam is measured at several angles using a photo-detector. If a spatially resolved measurement is desired (for example in structured light applications) then a camera can be used. By analysing the measured intensities, the Stokes parameters can be calculated. The rotating wave-plate method is relatively simple, low cost, and provides high accuracy. However, this method requires mechanical rotation, which can be challenging to implement in some applications. An alternative method uses DMDs to ``digitally'' measure the Stokes parameters of a beam, i.e. without mechanically moving parts (such as wave plates) \cite{Zhao2019,manthalkar2020}.

In this lab note, we present a simple yet novel static experimental arrangement and associated MATLAB code which we call a ``Stokes Camera''. The Stokes Camera enables full real-time measurement of spatial polarisation and thus spatially resolved Stokes parameters. This arrangement employs a polarisation sensitive camera coupled with a fixed quarter-wave plate and software processing, offering a simple yet effective solution for one-shot, digital polarization measurement.

To make this lab note as useful as possible to the reader, we provide an expanded background in Sec.~\ref{sec:background} on relevant polarisation theory as well as vector vortex beams (a form of structured light making use of all three degrees of freedom) which we use to demonstrate the operation of the Stokes Camera. Boradly speaking, we present this background in a ``top-down'' manner. We describe the experimental arrangement and required processing code in Sec.~\ref{sec:setupandcode}. In Sec.~\ref{sec:results} we provide several examples of output of the provided codes, which are open source and available on Github \cite{githubStokes}.

\section{Background}
\label{sec:background}

\subsection{The Polarisation Ellipse and the Poincar\'e Sphere}
In this section a summary of the various techniques and associated theory will be provided in a tutorial-like manner. For the readers convenience as a lab note we provide more detail than is strictly required to describe our ``Stokes Camera''. A comprehensive treatment of all the topics and theory herein would be formidable, and so, the interested reader is invited to read more in the references provided. We begin by writing the polarisation components of an electric field in the form of column vector as,
\begin{equation}
    \label{eq:jonesVector}
    \mathbf{E} = 
    \begin{bmatrix}
    E_x\\
    E_y
  \end{bmatrix},
\end{equation}
where $E_x$ is conventionally the complex amplitude (i.e. it has a magnitude and phase) of the field on the horizontal axis and $E_y$ on the vertical. In this form, this vector is known as the Jones vector and it can be conveniently transformed using various Jones matrices which describe different optical elements such as wave-plates and polarisers, thus allowing computation of the resulting polarisation of a beam, discussed further in Sec.~\ref{sec:jones}. As a side note, two polarisation vectors are orthogonal if the inner product between them is zero.

\begin{figure}[t]
    \centering
    \includegraphics[width=0.5\textwidth]{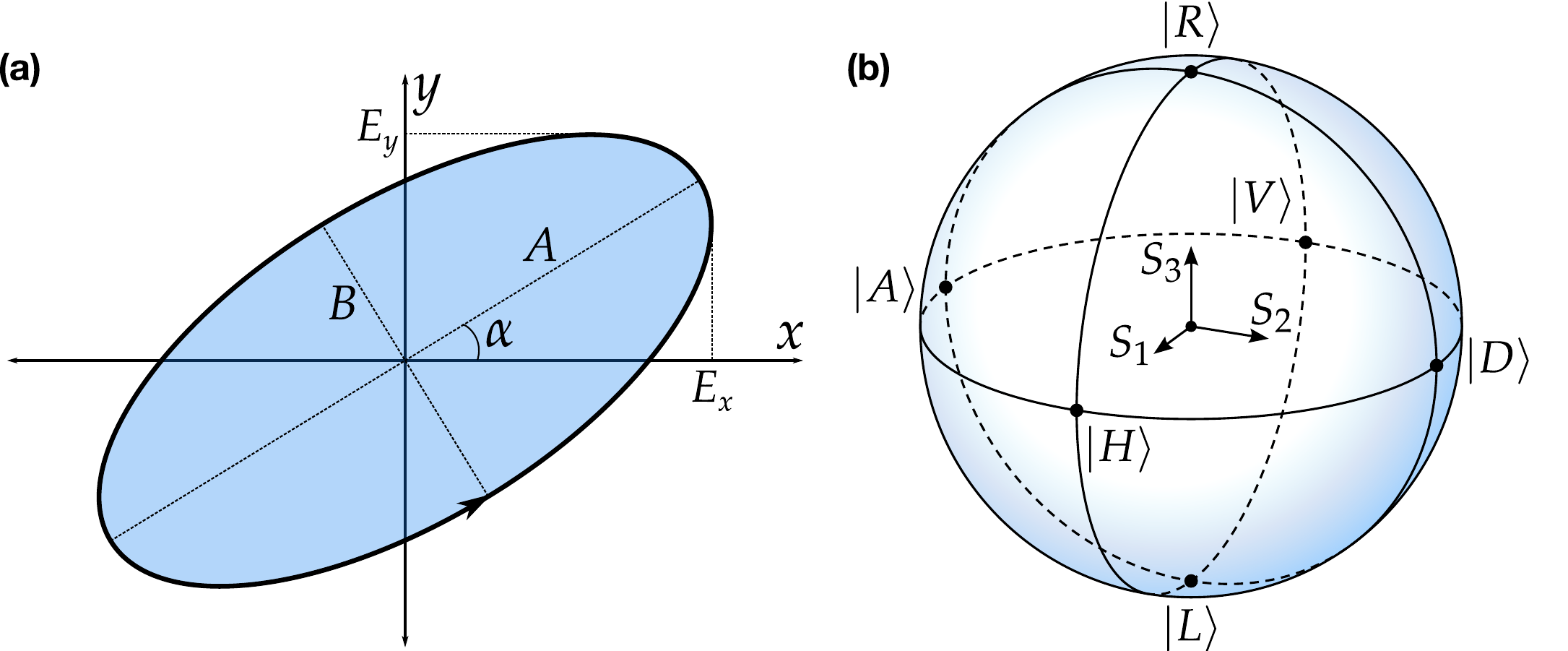}
    \caption{(a) The Poincar\'e sphere showing all the possible states of polarisation of light indicated in Dirac notation with relevant Stokes parameters ($S_{1,2,3}$) also visualised. An alternative visualisation is the polarisation ellipse in (b).}
    \label{fig:poincare}
\end{figure}

Visualisation of the polarisation of a field is highly convenient, as dealing purely with complex numbers is not always intuitive to the lay scientist or engineer. Since polarisation can be linear, circular or somewhere in between (elliptical), a useful representation of each polarisation state is through the polarisation ellipse, shown in Fig.~\ref{fig:poincare}~(a), which captures the essence of each polarisation state, the mathematical formalism of this will be derived below. A more useful representation relies on representing each of these ellipses in three dimensions using the surface of a unitary sphere, known as the Poincar\'e sphere, which shown in Fig.~\ref{fig:poincare}~(b), in honour of Henri Poincar\'e, the inventor of this representation \cite{Poincare1892}. In this representation, the right- and left-handed polarisation states are mapped to the north and south pole, respectively, the linear polarisation states along the equator, whereas all the elliptical states on the rest of the sphere. Representing each polarisation state through a specific polarisation can be specially useful to represent structured light beams with none homogeneous polarisation. More importantly such polarisation structure can be obtained from intensirty images, as will be explained below. These ellipses are often overlaid on the intensity profile of a beam that have spatial interest - we call this a Stokes plot for brevity.

To understand how polarisation can be represented either as an ellipse or on the surface of a sphere, it is convenient to represent each component of the electric field as a periodic function of the form
\begin{equation}
\begin{split}
\label{eq:EF}
    &E_x(t)=E_{0x}\cos(\tau+\delta_x)\\
   &E_y(t)=E_{0y}\cos(\tau+\delta_y)
\end{split}
\end{equation}
where $\tau=\omega t-kz$. In what follows we will perform a series of mathematical steps to derive the polarisation ellipse equation. To begin with, we can apply the trigonometric identity $\cos{(\theta_1+\theta_2)}=\cos\theta_1\cos\theta_2-\sin\theta_1\sin\theta_2$ to both expressions in Eq. \ref{eq:EF} and divide by $E_{0x}$ and $E_{0y}$, respectively, to obtain,
\begin{equation}
\begin{split}
    &\frac{E_x(t)}{E_{0x}}=\cos\tau\cos\delta_x-\sin\tau\sin\delta_x\\
    &\frac{E_y(t)}{E_{0y}}=\cos\tau\cos\delta_y-\sin\tau\sin\delta_y.
    \label{EqPol}
\end{split}
\end{equation}
After multiplying the first equation by $\sin\delta_y$ and the second by $\sin\delta_x$ and subtracting the second from the first we obtain
\begin{equation}
    \frac{E_x(t)}{E_{0x}}\sin\delta_y-\frac{E_y(t)}{E_{0y}}\sin\delta_x=\cos\tau\sin(\delta_x-\delta_y).
\end{equation}
Squaring both sides of the equation then yields
\begin{equation}
\begin{split}
         &\frac{E^2_x(t)}{E^2_{0x}}\sin^2\delta_y-2\frac{E_x(t)}{E_{0x}}\frac{E_y(t)}{E_{0y}}\sin\delta_x\sin\delta_y+\\
     &\frac{E^2_y(t)}{E^2_{0y}}\sin^2\delta_x=\cos^2\tau\sin^2(\delta_x-\delta_y).
\label{Eq1} 
\end{split}
\end{equation}

In a similar way, we can now multiply Eq.~\ref{EqPol} by $\cos\delta_x$ and $\cos\delta_y$, subtract both equations and square the result to obtain
\begin{equation}
\begin{split}
        &\frac{E^2_x(t)}{E^2_{0x}}\cos^2\delta_y-2\frac{E_x(t)}{E_{0x}}\frac{E_y(t)}{E_{0y}}\cos\delta_x\cos\delta_y+\\
        &\frac{E^2_y(t)}{E^2_{0y}}\cos^2\delta_x=\sin^2\tau\sin^2(\delta_x-\delta_y).
\label{Eq2}
\end{split}
\end{equation}
Finally, adding both equations yields the final result:
\begin{equation}
    \frac{E^2_x(t)}{E^2_{0x}}+\frac{E^2_y(t)}{E^2_{0y}}-2\frac{E_x(t)}{E_{0x}}\frac{E_y(t)}{E_{0y}}\cos\delta=\sin^2\delta,
    \label{Eq:PE}
\end{equation}
where, $\delta=\delta_x-\delta_y$. This equation is known as the polarisation ellipse because it has the general form of a rotated ellipse, as illustrate in Fig. \ref{fig:poincare}, where $A$ and $B$ represent the major and minor axis and $\alpha$ the angle of the major axis with respect to the component $E_x$ of the electric field. From this equation any polarisation state can be obtained, for example, for $\delta=0,\pi$ the above equation transforms into,
\begin{equation}
    E_x(t)=\pm \frac{E_{0x}(t)}{E_{0y}(t)} E_y(t),
\end{equation}
which is the equation of a line with slope $\pm E_{0x}(t)/E_{0y}(t)$, and therefore, it is associated to linear polarisation states. As another example, for $\delta=\pi/2,3\pi/2$, Eq.~\ref{Eq:PE} transforms into,
\begin{equation}
    \frac{E^2_x(t)}{E^2_{0x}}+\frac{E^2_y(t)}{E^2_{0y}}=1,
\end{equation}
which is the equation of a circle and therefore it describes circular polarisation states. 
The equation of the polarisation ellipse (Eq. \ref{Eq:PE}) can be transformed into its standard form (not rotated), from which we can obtain the relations,
\begin{equation}
\begin{split}
        &\tan(2\alpha)=\tan(2\phi)\cos\delta \qquad 0\leq\alpha\leq\pi \\
        &\sin(2\chi)=\sin(2\phi)\sin\delta, \qquad -\pi/4\leq\chi\leq\pi/4
        \label{StokesParameters}
\end{split}
\end{equation}
from which, any polarisation state can be obtained in terms of the the parameters $\alpha$ and $\chi$, which are associated to the orientation angle (including direction) and ellipticity angle of polarisation, respectively \cite{Goldstein2011}. Here, the parameter $\phi$ is related to the polarisation components of the electric field through the relation,
\begin{equation}
    \tan(\phi)=\frac{E_{0y}}{E_{0x}} \qquad 0\leq\phi\leq\pi/2.
\end{equation}
Notice that, for $-\pi/4\leq\chi\leq 0$ (see Eq. \ref{StokesParameters}) we have left handed circular polarisation, whereas for $0\leq\chi\leq\pi/4$ right handed. As a final comment, the major and minor axis of the polarisation ellipse, A and B, respectively, are related to the ellipticity angle $\chi$ through the relation,

\begin{equation}
    \tan(\chi)=\frac{\pm B}{A}.
\end{equation}

\subsection{The Stokes parameters}
\label{sec:stokesparams}

In the polarisation ellipse equation, $E_x$ and $E_y$ are both implicitly dependent on time, hence, it predicts a temporal evolution of polarisation, but the fast oscillations of the electromagnetic field does not make possible an experimental observation. Hence, in order to observe the effect of polarisation, we must take an average over the time of observation of the previous equation, 
\begin{equation}
    \frac{\langle E^2_x(t) \rangle}{E^2_{0x}}+\frac{\langle E^2_y(t) \rangle }{E^2_{0y}}-2\frac{\langle E_x(t)E_y(t) \rangle}{E_{0xE_{0y}}}\cos\delta=\sin^2\delta,
\label{Eq3}
\end{equation}
Using the time average for each term, whic are given by,
\begin{equation}
\begin{split}
    &\langle E^2_x(t) \rangle=\frac{E^2_{0x}}{2}\\
    &\langle E^2_y(t) \rangle=\frac{E^2_{0y}}{2}\\
    &\langle E_x(t)E_y(t) \rangle=\frac{1}{2} E_{0x}E_{0y},
\label{Eq4}
\end{split}
\end{equation}
the previous equation can be written as
\begin{equation}\begin{split}
    &(E^2_{0x}+E^2_{0y})^2-(E^2_{0x}-E^2_{0y})^2-(2E_{0x}E_{0y}\cos\delta)^2=\\
    &(2E_{0x}E_{0y}\sin\delta)^2,
    \label{Stokes1}    
\end{split}
\end{equation}
At this point, it is useful to introduce the following definitions,
\begin{equation}
\begin{split}
        &S_0=E^2_{0x}+E^2_{0y}, \hspace{10mm} S_2=2E_{0x}E_{0y}\cos\delta\\
        &S_1=E^2_{0x}-E^2_{0y}, \hspace{8mm} S_3=2E_{0x}E_{0y}\sin\delta ,
        \label{StokesParameters2}
\end{split}
\end{equation}
which are known as the Stokes parameters. In terms of these parameters Eq.~\ref{Stokes1} can be written as,
\begin{equation}
    S_0^2=S_1^2+S_2^2+S_3^2.
\end{equation}
The Stokes parameters can also be written using a complex representation for the electric field as, 
\begin{equation}
\begin{split}
        &S_0=E_xE^*_x+E_yE^*_y, \hspace{10mm} S_2=E_xE^*_y+E_yE^*_x\\
        &S_1=E_xE^*_x-E_yE^*_y, \hspace{8mm} S_3=i(E_xE^*_y-E_yE^*_x),
        \label{StokesComplex}
\end{split}
\end{equation}
where, $E^*_x$ and $E^*_y$ represent the complex conjugated of $E_x$ and $E_y$, respectively. A more useful representation based only on intensity measurements will be derived below, since this allows a reconstruction of any polarisation state.

In summary, $S_0$ represents the total intensity of the light field, while $S_1$ describes the amount of linear polarisation in the horizontal and vertical directions. $S_2$ describes the amount of diagonal polarisation in the $45\degree$ and $135\degree$ directions, while $S_3$ describes the amount of circular polarisation in the left and right directions.


\subsection{Stokes Polarimetry}
\label{sec:stokes}

The primary goal of Stokes polarimetry is to measure the polarisation structure of an optical field (see Sec.~\ref{sec:stokesparams}), and thus describe the intensity and polarisation properties of the light field. 


For the sake of clarity, we will explain how the Stokes parameters can be obtained from intensity measurements performed with a polariser and a phase retarder. To this end, we write the components of the electric field using complex notation, namely,
\begin{equation}
    \begin{split}
        &E_x(t)=E_{0x}\exp{(i\delta_x)}\exp{(i\omega t)}\\
        &E_y(t)=E_{0y}\exp{(i\delta_y)}\exp{(i\omega t)}
    \end{split}
\end{equation}

As a first step, the beam is sent through a Quarter-Wave Plate (QWP), which produces a phase delay $\theta_1$ between both components. As result, the components of the new electric field take the form
\begin{equation}
    \begin{split}
        &E'_x(t)=E_{x}(t)\exp{(i\theta_1/2)}\\
        &E'_y(t)=E_{y}(t)\exp{(-i\theta_1/2)}
    \end{split}
\end{equation}
The beam that exits the QWP is then sent through a Linear Polariser (LP), which transmits light only along the transmission axis of the same. Mathematically, the total light transmitted through the LP, considering its transmission axis at an angle $\theta_2$ from the horizontal, is described by,
\begin{equation}
\begin{split}
   & E_f(\theta_1,\theta_2)=\\
    &E_{x}(t)\exp{(i\theta_1/2)}\cos\theta_2+E_{y}(t)\exp{(-i\theta_1/2)}\sin\theta_2.
\end{split}
\end{equation}
Since we can only measure intensities experimentally, we have to compute the intensity through the relation  $E_f(\theta_1,\theta_2)\cdot E^*_f(\theta_1,\theta_2)$,
\begin{widetext}
    \begin{equation}
I(\theta_1,\theta_2)=E_{x}E^*_{x}\cos^2\theta_2+E_{y}E^*_{y}\sin^2\theta_2+E_yE^*_{x}\exp{(-i\theta_1)}\sin\theta_2\cos\theta_2+E^*_yE_{x}\exp{(i\theta_1)}\sin\theta_2\cos\theta_2,
\end{equation}
\end{widetext}

which can be transformed, using trigonometric identities into,
\begin{equation}
\begin{split}
I(\theta_1,\theta_2)&=\frac{1}{2}[E_{x}E^*_{x}+E_{y}E^*_{y}+(E_{x}E^*_{x}-E_{y}E^*_{y})\cos2\theta_2\\
&+(E_yE^*_{x}+E^*_yE_{x})\cos\theta_1\sin2\theta_2+\\
&i(E_yE^*_{x}-E^*_yE_{x})\sin\theta_1\sin2\theta_2],
\label{Intensity}
\end{split}
\end{equation}
We can now use Eq.~\ref{StokesComplex} to rewrite Eq.~\ref{Intensity} in terms of the Stokes parameters as,
\begin{equation}
\begin{split}
  & I(\theta_1,\theta_2)=\\
&\frac{1}{2}[S_0+S_1\cos2\theta_2+S_2\cos\theta_1\sin2\theta_2+S_3\sin\theta_1\sin2\theta_2].
\end{split}
\end{equation}
Specific settings for the angles of the QWP and LP, $\theta_1$ and $\theta_2$, respectively, give the relations,
\begin{equation}
    \begin{split}
        &I(0,0)=\frac{1}{2}(S_0+S_1)\qquad I(0,90\degree)=\frac{1}{2}(S_0+S_1)\\
        &I(0,45\degree)=\frac{1}{2}(S_0+S_1)\qquad I(90\degree,45\degree)=\frac{1}{2}(S_0+S_1),
    \end{split}
\end{equation}
from which we can express the Stokes parameters in terms of intensity values as,
\begin{equation}
    \begin{split}
        &S_0=I(0,0)+I(0,90\degree),\\
        &S_1=I(0,0)-I(0,90\degree),\\
         &S_2=2I(0,45\degree)-[I(0,0)+I(0,90\degree)]\\
        &=2I(0,45\degree)-S_0,\\
        & S_3=2I(90\degree,45\degree)-[I(0,0)+I(0,90\degree)\\
        &=2I(90\degree,45\degree)-S_0.
    \end{split}
\end{equation}
The above equations can be interpreted as follows, the parameter $S_0$, the total intensity of the beam, can be measured as the addition of the intensity when both, the QWP and the LP are set to $0\degree$, which is the intensity corresponding to the horizontal polarisation $I_H$, plus the intensity when the QWP is set to $0\degree,$ and the LP to $90\degree$, which corresponds to the vertical polarisation component $I_v$. The Stokes parameter $S_1$ is simple the subtraction of the horizontal and vertical polarisation components. The intensity $I(0,45\degree)$ corresponds to diagonal polarisation $I_D$, whereas the intensity $I(90\degree,45\degree)$ to right circular polarisation $I_R$. Hence, the Stokes parameters can be written simply as,
\begin{equation}
    \label{eq:stokesParams}
    \begin{split}
        &S_0 = I_H + I_V , \\
        &S_1 = I_H - I_V, \\
        &S_2 = 2I_D - S_0, \\
        &S_3 = 2I_R - S_0, \\
    \end{split}
\end{equation}
Using that the Stokes parameter $S_0$ can also be written as the addition of diagonal $I_D$ plus antidiagonal $I_A$ polarisation, {\it i.e.} $S_0=I_D+I_A$ and similarly as the addition of right- plus left-handed circular polarisation, {\it i.e.} $S_0=I_R+I_L$, the above equations can also be written as,
%
\begin{equation}
    \label{eq:stokesParams6}
    \begin{split}
        &S_0 = I_H + I_V  \\
        &S_1 = I_H - I_V, \\
        &S_2 = I_D - I_A, \\
        &S_3 = I_R - I_L. \\
    \end{split}
\end{equation}
Table \ref{tab:Stokes} summarises the different configurations of the QWP and LP to measure the required intensities to compute the four Stokes parameters.
\begin{table}
\centering
	\caption{\label{tab:Stokes}Configuration of the QWP and LP to measure the required intensities to compute the four Stokes parameters
 }
	\begin{tabular}{rcc}
    	\hline
     \centering
		Intensity & QWP & LP  \\ 
        \hline
		$I_H$ & $0\degree$ & $0\degree$ \\ 
		$I_V$ & $0\degree$ & $90\degree$  \\ 
		$I_D$ & $0\degree$ & $45\degree$  \\ 
		$I_R$ & $90\degree$ & $45\degree$  \\ 
        \hline
	\end{tabular} 
\end{table}

The spatial distribution of the polarisation properties can be obtained by measuring these parameters at multiple points across the beam using a camera or other imaging device, which we call a Stokes plot. This distribution can be plotted as a polarisation ellipse, which provides an intuitive way to visualise the polarisation properties of the light field in two dimensions. The major and minor axes of the ellipse represent the polarisation directions, while the eccentricity of the ellipse represents the ellipticity or degree of circular polarisation.

\subsection{Jones Calculus}
\label{sec:jones}

While the theory of Jones calculus and Vector Vortex Modes (Sections~\ref{sec:jones} and \ref{sec:vvm}) is not strictly necessary to understand the work presented in this lab note, we present it nonetheless as we believe it is very useful for the interested reader to form a well-rounded understanding of polarisation. 

It is convenient to express the polarisation states as a column vector and optical elements as a 4-entry matrix. This ``Jones calculus'' makes it easy to obtain the polarisation state of a polarised beam after traversing a single or a set of optical elements. In terms of a normalised Jones vector (where the intensity is one) we rewrite Eq.~\ref{eq:jonesVector} as
\vspace{-0.5cm}
\begin{subequations}
\label{eq:jonesVector2}

\begin{equation}
\label{eq:jonesELinear}
    \mathbf{E}_\mathrm{linear} =
    \begin{bmatrix}
    \cos\theta\\
    \sin\theta
    \end{bmatrix}, 
\end{equation}

\begin{equation}
\label{eq:jonesECircular}
    \mathbf{E}_\mathrm{circular} = \frac{1}{\sqrt{2}}
    \begin{bmatrix}
    1\\
    \pm i
    \end{bmatrix}.
\end{equation}

\end{subequations}
\noindent Note that we use $\theta$ to indicate the rotation of the linear polarisation from the $x$-axis (horizontal), and $i$ is the complex number. For left-handed circular polarisation the helicity is positive. For quick reference, and in Dirac (``ket'') notation for brevity, these six standard polarisation's are given by 
\vspace{-0.5cm}
\begin{subequations}
\label{eq:jonesDirectional}

\begin{equation}
  \ket{H}
  =
  \begin{bmatrix}
    1\\
    0
  \end{bmatrix},
\end{equation}

\begin{equation}
  \ket{V}
  =
  \begin{bmatrix}
    0\\
    1
  \end{bmatrix},
\end{equation} 

\begin{equation}
  \ket{D}
  =
  \frac{1}{\sqrt{2}}
  \begin{bmatrix}
    1\\
    1
  \end{bmatrix},
\end{equation}

\begin{equation}
  \ket{A}
  =
  \frac{1}{\sqrt{2}}
  \begin{bmatrix}
    1\\
    -1
  \end{bmatrix},
\end{equation}

\begin{equation}
  \ket{R}
  =
  \frac{1}{\sqrt{2}}
  \begin{bmatrix}
    1\\
    -i
  \end{bmatrix},
\end{equation}

\begin{equation}
  \ket{L}
  =
  \frac{1}{\sqrt{2}}
  \begin{bmatrix}
    1\\
    i
  \end{bmatrix}.
\end{equation}

\end{subequations}

\noindent Of relevance to this lab note are quarter-wave plates (QWP), half-wave plates (HWP), linear polarisers and Q-plates \cite{Marrucci2006}, with their respective Jones matrices given by
\begin{subequations}
\label{eq:jonesVector3}
\begin{equation}
\label{eq:jonesQWP}
    \mathrm{QWP}: \frac{1}{\sqrt{2}}
    \begin{bmatrix}
    1+i\cos2\theta & i\sin2\theta\\
    i\sin2\theta & 1 - i\cos2\theta
    \end{bmatrix}, 
\end{equation}

\begin{equation}
\label{eq:jonesHWP}
    \mathrm{HWP}:
    \begin{bmatrix}
    \cos2\theta & \sin2\theta\\
    \sin2\theta & -\cos2\theta
    \end{bmatrix}, 
\end{equation}
    
\begin{equation}
\label{eq:jonesLP}
    \mathrm{Linear~Polariser}:
    \begin{bmatrix}
    \cos^2\theta & \sin\theta \cos\theta \\
    \sin\theta \cos\theta & \sin^2\theta
    \end{bmatrix}.   
\end{equation}

\begin{equation}
\label{eq:jonesQP}
    \mathrm{Q-plate}:
    \begin{bmatrix}
    \cos2q\psi & \sin2q\psi \\
    \sin2q\psi & -\cos2q\psi
    \end{bmatrix}, 
\end{equation}
\end{subequations}
\noindent In the special case of the Q-plate, $\psi$ is a spatially resolved azimuthal angle which describes the azimuthally varying polarisation (i.e. vector beam) that is produced by the Q-plate. 

In order to calculate the final polarisation of a laser beam that has propagated through several elements, $\mathbf{E}^{'}$, we simply write the matrices for the various elements in reverse propagation order (i.e. the last element is written first), with the incoming beam Jones vector last, and then perform the multiplication. For example, assuming a diagonally polarised source is passed through a horizontal polariser and then a quarter-wave plate at an angle of $45\degree$, the calculation is as follows
\begin{widetext}
\begin{equation}
    \begin{bmatrix}
      E_x^{'}\\
      E_y^{'}
    \end{bmatrix}
    = 
    \underbracket{\frac{1}{\sqrt{2}}
    \begin{bmatrix}
    1+i\cos2\theta & i\sin2\theta\\
    i\sin2\theta & 1 - i\cos2\theta
    \end{bmatrix}}_{\mathrm{QWP,}~\theta=45\degree}
    \underbracket{
    \begin{bmatrix}
    \cos^2\theta & \sin\theta \cos\theta \\
    \sin\theta \cos\theta & \sin^2\theta
    \end{bmatrix}}_{\mathrm{Polariser,}~\theta=0\degree}
    \underbracket{
    \frac{1}{\sqrt{2}}
    \begin{bmatrix}
    1\\
    1
   \end{bmatrix}}_{\mathrm{Beam,}~\theta=45\degree}
   = \frac{1}{2}
   \begin{bmatrix}
    1\\
    i
   \end{bmatrix},
\end{equation}
\end{widetext}
\noindent which is, as we would expect, a left-polarised beam with reduced intensity.

\subsection{Vector Vortex Modes}
\label{sec:vvm}

While the focus of this lab note is not on vector modes or $q$-plates, we will briefly discuss few examples of vector modes, as they are the most general state on polarisation, non-homogeneous, and have attracted a great amount of attention in fields such as optical communications, optical tweezers optical metrology, amongst others \cite{Rosales2018Review,Yuanjietweezers2021,Ndagano2018}. Given the interest in vector beams, we will use cylindrical vector vortex (CVV) modes to illustrate the operation of the device. CVV modes have a donut-like intensity profile, which are generated from the weighted superposition of light beams with azimuthally varying spatial phase in opposite direction, which carry orbital angular momentum (OAM) \cite{Cox:16}. Crucially, they also have an azimuthally varying polarisation, the direction of which depends on the type of mode as described below. 

The generation of both scalar and CVV modes can be accomplished using a $q$-plate in combination with quarter- and half- wave plates according to the unitary operation $\hat{Q}$ of a $q$-plate\cite{Marrucci2011}:
%
\begin{subequations}
\label{eq:qTransform}
\begin{equation}
\hat{Q}
\begin{bmatrix}
    \ket{R}\\
    \ket{L} 
  \end{bmatrix}
  =
  \begin{bmatrix}
    \ket{+2q}\ket{L}\\
    \ket{-2q}\ket{R} 
  \end{bmatrix},
\end{equation}
\begin{equation}
\hat{Q}
\begin{bmatrix}
    \ket{H}\\
    \ket{V} 
  \end{bmatrix}
  =
  \begin{bmatrix}
    \ket{+2q}\ket{R} + \ket{-2q}\ket{L}\\
    \ket{+2q}\ket{R} - \ket{-2q}\ket{L}
  \end{bmatrix},
\end{equation}  

\end{subequations}
\noindent where $q$ is the charge of the topological singularity of the $q$-plate (in this work $q=1/2$), related to the OAM topological charge, $\ell=2q$. When the input beam is circularly polarized, the output is a scalar beam of the opposite polarisation embedded with an OAM of $\pm\ell$. Alternatively, when the input is linearly polarised the output is a CVV in which polarisation and OAM are coupled. As such, there are four main types of CVV mode with specific names as a legacy from radio frequency wave-guide theory, as follows:
%
\begin{subequations}

\begin{equation}
\ket{TM} = \frac{1}{\sqrt{2}}(\ket{\ell}\ket{R} + \ket{-\ell}\ket{L}),
\end{equation}

\begin{equation}
\ket{TE} = \frac{1}{\sqrt{2}}(\ket{\ell}\ket{R} - \ket{-\ell}\ket{L}),
\end{equation}

\begin{equation}
\ket{HE^e} = \frac{1}{\sqrt{2}}(\ket{\ell}\ket{L} + \ket{-\ell}\ket{R}),
\end{equation}

\begin{equation}
\ket{HE^o} = \frac{1}{\sqrt{2}}(\ket{\ell}\ket{L} - \ket{-\ell}\ket{R}).
\end{equation}

\label{eq:vector}
\end{subequations}

\subsection{Polarisation Cameras}
\label{sec:camera}

The polarisation camera used in this work makes use of the Sony IMX250 sensor \cite{Yamazaki2016,SonySemiconductorSolutionsGroup}. The sensor is relatively large, with an 11.1~mm diagonal (2/3 inch), a resolution of 2464$\times$2056 3.45~$\mu$m pixels and a maximum frame rate of about 163 frames per second at full resolution (although this would be dependant on the specific camera, drivers, region of interest, etc.). This sensor is readily available in several popular brands of camera.

Rather than red-green-blue colour filters in front of the pixels in the camera, a special polarising wire grid is used which is situated between an array of micro-lenses and the photodiode layer. The polarisation filter repeats in a $2\times2$ grid with linear polarisation's [H D; A V], shown in Fig~\ref{fig:camera}~(c). According to the sensor datasheet, the extinction ratios at 532~nm and 633~nm are approximately 300:1 and 130:1, respectively. For comparison, a typical Thorlabs polariser has an extinction ratio of about 1000:1, so very high precision measurements are unfortunately not possible. Perhaps newer versions of the sensor will improve on this, however, for an 8~bit image the error due to the extinction ratio is on the same order of magnitude as the quantisation error.

\section{Experimental Setup and Analysis Code}
\label{sec:setupandcode}

\subsection{Experimental Setup}

\begin{figure}[h]
    \centering
    \includegraphics[width=0.45\textwidth]{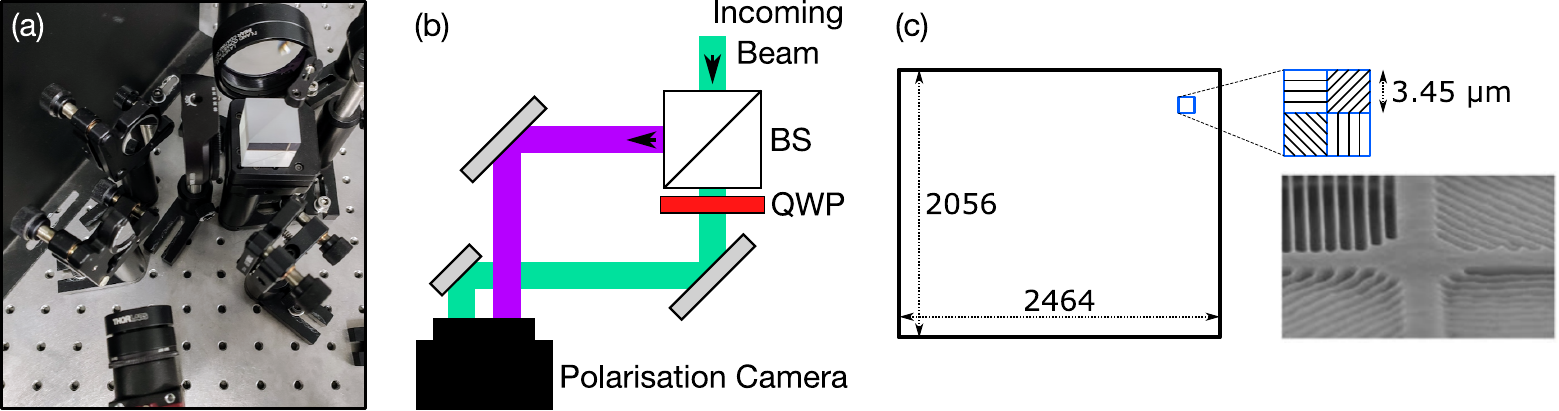}
    \caption{The structure of the polarisation camera pixels with an inset photograph of the wire-grid (credit \cite{Yamazaki2016}).}
    \label{fig:camera}
\end{figure}

Since the camera is only able to measure linear polarisation states, we must use a quarter wave plate to transform the circular components of the incoming beam into linear polarisations. We then measure the original beam as well as the transformed beam simultaneously, and use some software processing to extract the relevant information.

A photo and drawing of the setup is shown in Fig~\ref{fig:camera}~(a) and (b), respectively. The setup is quite tight, but easy to align. A 50:50 beam splitter (not polarising) creates the aforementioned split, and several mirrors are used to direct the two resulting beams onto halves of the camera sensor. It is important that these two beams travel in parallel. Any tilt on one of the beams with respect to the other will result in some error. As such, do not be tempted to direct the beam from the beam splitter directly to the camera. One should also be aware that a mirror flips the circular polarisation of a beam, and so it is important to have an even number of mirrors after the quarter wave plate.

\subsection{Analysis Code}

The code to calculate the Stokes images, representing $S_0$ to $S_3$, is straightforward but not trivial and is provided on GitHub \cite{githubStokes}. Users who make enhancements and fixes to the code are encouraged to submit pull requests for the good of the community.

In essence, on the first run, the code performs a cross-correlation between the two halves of the camera image. This is cached for subsequent runs as it takes a few minutes, depending on the resolution of the camera image or region of interest. After this is done, the measurements are robust to misalignments of the incoming beam, as long as the components of the Stokes camera (the beam splitter, wave plate, and mirrors) are not moved. The offset between the two halves is found using this cross-correlation and is used to digitally align them for the next step. For each frame, the different polarisation components are separated (see Sec.~\ref{sec:camera}), offset accordingly, and the arithmetic (see Eq.~\ref{eq:stokesParams6}) is performed to calculate the four Stokes images.

The code is designed to run continuously, as a camera preview window. We recommend running this preview in a dedicated MATLAB instance so that the window can remain open while other commands are executed. In addition, if the user presses `p' while the window is in focus, all of the latest frames are stored to a PNG image and a MATLAB file for later processing. The GitHub repository contains additional information and tips, which are expected to accumulate over time in the spirit of open source.

In addition, we provide a code that can be used separately to plot so-called Stokes plots. These plots show $S_0$ (the intensity of the beam) with a number of polarisation ellipses overlaid, representing the state of polarisation at those points of the beam.

\section{Results and Discussion}
\label{sec:results}

\begin{figure}[tb]
    \centering
    \includegraphics[width=0.45\textwidth]{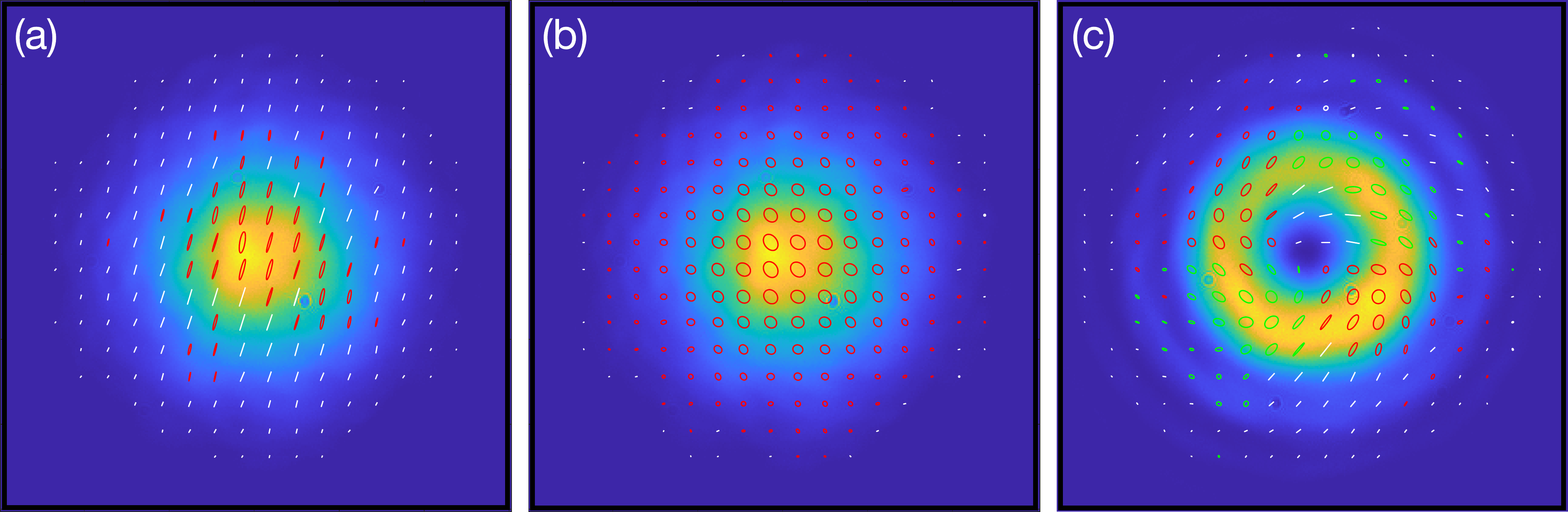}
    \caption{Stokes plots of example beams (from left to right: approximately vertically polarised, circularly polarised and a vector vortex mode) measured with the stokes camera and plotted showing the polarisation ellipses over the beam which are generated using the four stokes images. The polarisation of these beams was controlled using two fibre paddles and in (c) a Q-plate.}
    \label{fig:beams}
\end{figure}

We tested the Stokes Camera using a 520~nm, green single-mode fibre-coupled laser diode. Two fibre paddles were used to control the state of polarisation of the output Gaussian-like beam. Using the Stokes camera preview window, we were able to see the polarisation of the beam in real-time, much like commercially available polarimeters. Two example Gaussian beams are shown in Fig.~\ref{fig:beams} (a) and (b).

For our second test, we inserted a Q-plate in the path of the beam \cite{Rosales2018Review,Yuanjietweezers2021,Ndagano2018,Cox:16}. Figure~\ref{fig:beams}~(c) shows one example of the resulting vector vortex mode.

The Stokes camera works well, is extremely convenient compared to manual methods, and is capable of real-time observation and measurement. With a limited region of interest and a GigE version of the polarisation camera, we achieved 11~frames per second. This could be improved by using a USB3 camera, which has a faster inherent frame rate (5~Gbps versus 1~Gbps). The computational requirements of the code have been optimised significantly (for example, caching the cross-correlation), but are still formidable. We believe there is further room for optimisation, but a Python or C++ implementation would likely enable faster frame rates. 

If fast measurements are required, then a video of the camera output (at hundreds of frames per second, with the right configuration and region of interest) could be captured without processing, and simply processed offline using our code.

The dynamic range of our Stokes Camera is the limiting factor compared with commercial polarimeters. Since an 8 bit image is used (i.e. [0; 255]), the dynamic range is approximately 48~dB. Using a higher bit depth (for example, 10 bit) on the camera is unnecessary, as the extinction ratio is only 1:300, which fundamentally only provides about a 1~dB improvement at the expense of frame rate. In practice, we believe the measurements using this system are likely to be superior to manually operated Stokes measurement systems often used in experiments.

\section{Summary}
This paper presents a self-contained tutorial aimed at simplifying the task of real-time reconstruction of the transverse polarisation distribution of an optical beam using a polarisation camera. One aim of this tutorial is to popularise Stokes polarimetry amongst a broader audience, encompassing both researchers and laboratory classes. To this end, we start by clarifying basic concepts from a theoretical standpoint, including the polarisation ellipse, the Poincaré sphere, and the Stokes parameters. Furthermore, we outline how a specific polarisation state might change after passing through single or multiple optical polarisation components, such as linear polarisers, wave retarders, and q-plates, using straightforward matrix multiplication. We also provide a brief overview of vector beams—general light states characterised by a non-homogeneous polarisation distribution. Our focus on vector beams is twofold: they have recently gained considerable attention in the field, and they exemplify the effectiveness of polarisation cameras in reconstructing the transverse polarisation state of any optical beam, regardless of its polarisation complexity. Crucially, real-time reconstruction of the transverse polarisation distribution of vector beams has potential applications in optical metrology, especially for characterising dynamic systems sensitive to polarisation. This highlights the importance of timely and accurate methods for such beam polarisation distribution reconstructions. The early sections of the tutorial are supported with experimental results, where we detail the use of the polarisation camera and describe the necessary experimental setup to simultaneously capture all intensity images needed to compute the Stokes parameters. For hands-on application, we have included a set of MATLAB scripts to streamline the polarisation reconstruction process.

\section*{Funding}
NRF TTK2204011621; NRF NLCRPP221109667491; Wits Friedel Sellschop Award


\section*{Disclosures}
The author declares no conflicts of interest.

\section*{Data availability} Data underlying the results presented in this paper are not publicly available at this time but may be obtained from the authors upon reasonable request.


%

\end{document}